\documentstyle[12pt,epsfig]{article}
\title{\hspace{10cm} {\large Budker INP 97--54} \\ 
\vspace{0.5cm} 
 {\bf Progress in Photon Colliders}    
\footnote{Talk at the Intern. Conference ``Photon 97'', Egmond ann
Zee, The Netherland, May 10--15, 1997}}
\author{Valery Telnov \\
{\small\it Institute of Nuclear Physics,
630090, Novosibirsk, Russia}} 
\date{}                            
\begin{document}
\newcommand{\EP}{\mbox{e$^+$}}
\newcommand{\EM}{\mbox{e$^-$}}
\newcommand{\EPEM}{\mbox{e$^+$e$^-$}}
\newcommand{\EMEM}{\mbox{e$^-$e$^-$}}
\newcommand{\GG}{\mbox{$\gamma\gamma$}}
\newcommand{\GE}{\mbox{$\gamma$e}}
\newcommand{\GP}{\mbox{$\gamma$e$^+$}}
\newcommand{\TEV}{\mbox{TeV}}
\newcommand{\GEV}{\mbox{GeV}}
\newcommand{\LGG}{\mbox{$L_{\gamma\gamma}$}}
\newcommand{\EV}{\mbox{eV}}
\newcommand{\CM}{\mbox{cm}}
\newcommand{\MM}{\mbox{mm}}
\newcommand{\NM}{\mbox{nm}}
\newcommand{\MKM}{\mbox{$\mu$m}}
\newcommand{\SEC}{\mbox{s}}
\newcommand{\CMS}{\mbox{cm$^{-2}$s$^{-1}$}}
\newcommand{\MRAD}{\mbox{mrad}}
\newcommand{\IND}{\hspace*{\parindent}}
\newcommand{\E}{\mbox{$\epsilon$}}
\newcommand{\EN}{\mbox{$\epsilon_n$}}
\newcommand{\EI}{\mbox{$\epsilon_i$}}
\newcommand{\ENI}{\mbox{$\epsilon_{ni}$}}
\newcommand{\ENX}{\mbox{$\epsilon_{nx}$}}
\newcommand{\ENY}{\mbox{$\epsilon_{ny}$}}
\newcommand{\EX}{\mbox{$\epsilon_x$}}
\newcommand{\EY}{\mbox{$\epsilon_y$}}
\newcommand{\BI}{\mbox{$\beta_i$}}
\newcommand{\BX}{\mbox{$\beta_x$}}
\newcommand{\BY}{\mbox{$\beta_y$}}
\newcommand{\SX}{\mbox{$\sigma_x$}}
\newcommand{\SY}{\mbox{$\sigma_y$}}
\newcommand{\SZ}{\mbox{$\sigma_z$}}
\newcommand{\SI}{\mbox{$\sigma_i$}}
\newcommand{\SIP}{\mbox{$\sigma_i^{\prime}$}}
\maketitle
\begin{abstract}
Last two years were very important in history of a photon
  colliders. This option is included now in conceptual
  design reports of the NLC, JLC and TESLA/SBLC projects. All the
  designs foresee two interaction regions: one for \EPEM\ and the
  second for \GG, \GE\ and \EMEM\ collisions.  This paper is focused on
  three aspects: 1) arguments for photon colliders; 2) parameters of
  current projects; 3) ultimate luminosities and energies, new ideas.
  Recent studies have shown that the main collision effect - coherent
  pair creation - is suppressed at photon colliders with the energy
  (2E $<$ 2 TeV) due to the beam repulsion, and one can achieve, in
  principle, the \GG\ luminosity exceeding $10^{35}\; \CMS\ $.  The
  required electron beams with very small emittances can be obtained,
  for example, using a laser cooling of electron beams. This new
  method requires a laser with a power by one order of magnitude
  higher than that required for the ``conversion'' of electrons to
  photons. Such lasers are not available today, but hopefully they
  will appear by the time when linear colliders will be built.  High energy
  \GG, \GE\ colliders with the luminosity comparable to that in \EPEM\
  collisions are beyond the competition in study of many phenomena of
  particle physics.
\end{abstract}

\section{Introduction}

\begin{figure}[!hbp]
\centering
\epsfig{file=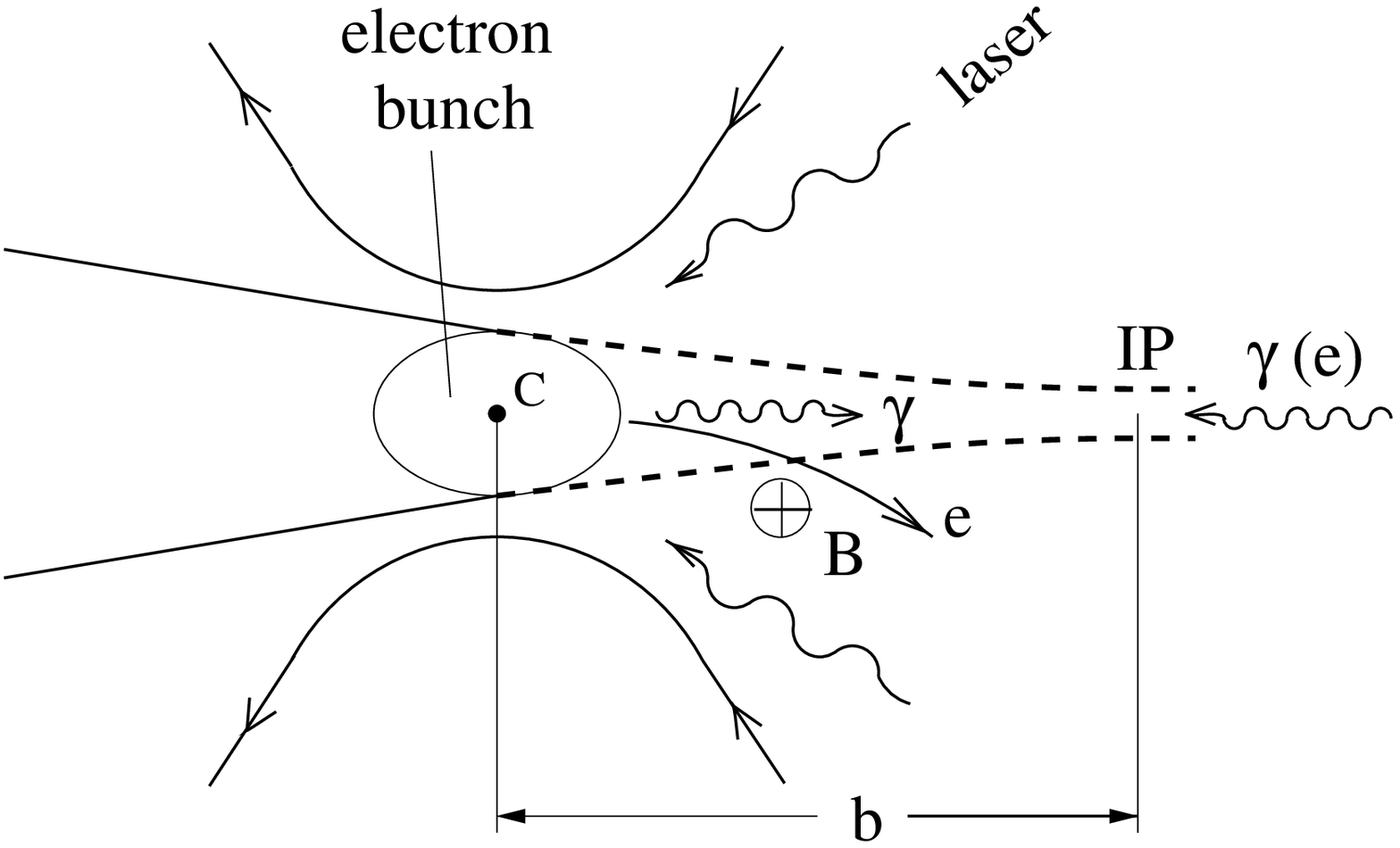, height=8.5cm,angle=-90}
\caption{Scheme of  \GG; \GE\ collider.}
\label{gensch}
\end{figure}

\IND\ Let me remind briefly the basic scheme of a photon
collider~\cite{GKST81},\cite{GKST83}, see fig.\ref{gensch}. Two electron
beams after the final focus system are traveling toward the
interaction point (IP).  At a distance of about 0.1--1 cm upstream from the
IP, at the conversion point (C), the laser beam is focused and Compton
backscattered by  electrons, resulting in the high energy beam of
photons.  With reasonable laser parameters one can ``convert'' most of
electrons into high energy photons. The photon beam follows the original
electron direction of motion with a small angular spread of order
$1/\gamma $, arriving at the IP in a tight focus, where it collides
with the similar opposing high energy photon beam or with an electron
beam. The photon spot size at the IP may be almost equal to that of
electrons at IP and therefore, the luminosity of \GG, \GE\ collisions
will be of the same order of magnitude as the ``geometric'' luminosity of basic
$ee$ beams.  The detailed description of photon
colliders properties can be found in
refs~\cite{GKST83}--\cite{TEL95} and
in the Berkeley Workshop Proceedings~\cite{BERK}.

  Now, this option is included into the Conceptual Design Reports of
the NLC~\cite{NLC}, TESLA--SBLC~\cite{TESLA} and JLC~\cite{JLC}. All
these linear collider projects foresee the second interaction region
for \GG,\GE\ collisions. This is quite a success but for final
decisions it is necessary to have very clear justification that photon
colliders are realistic and can substantially add to a discovery
potential of linear colliders. Below are some arguments for photon
colliders.

1.~Some phenomena can be studied at photon colliders better than
  anywhere, for example, the measurement of \GG\ $\to$ Higgs width, which
  is sensitive to all heavy charged particles; study of the vertex
$\gamma\gamma$WW.

2.~Cross sections for the pair production of charged scalar, leptons and
  top in \GG\ collisions are larger than those in \EPEM\ collisions by a
  factor of 5; for WW production this factor is even larger: 10--20.

3.~In \GE\ collisions charged supersymmetric particles with masses
higher than in \EPEM\ collisions  can be produced
(heavy charged particle plus light neutral).

4.~The luminosity of photon colliders (in the high energy part of
luminosity spectrum) with electron beam parameters considered in the
present designs will be about $10^{33}\; \CMS$ or by a factor 5 smaller
than $L_{\EPEM}$.  But the absence of collisions effects at 0.1 -- 1 TeV
photon colliders allows to reach \LGG\ up to $10^{35}$ \CMS\ using electron
beams with very low emittances. High luminosity photon colliders can
provide two orders high production rate of WW pair and other charged
particles (see item 2).

5.~Obtaining of the ultimately high luminosities requires the development of
new techniques, such as the laser cooling of electron
beams~\cite{TSB1}. However, linear colliders will appear (may be) only
in one decade and will work next two decades.  The upgrading of the
luminosity requires the injection part modification only; it may
be a separate injector for a photon collider, merging of many low
emittance RF-photoguns (with or without laser cooling) is one of
possible variants.

6.~Linear colliders are very expensive facilities and their potential
should be used in the best way. Two detectors (one for \EPEM\ and the
other for \GG\ and \GE) can give much more results than the simple doubling of
statistics in \EPEM\ collisions with one detector . 

7.~Development of X-ray FEL lasers based on linear colliders (which
 are now under way) will favour the work on FEL required for photon
 colliders.
                                                                      

\section{Physics potential, requirements to \GG\ luminosity }
  The physics in \GG, \GE\ colliders is very
rich. The total number of papers devoted to the physics at photon colliders
approaches to one thousand. Some examples are given in the
introduction. Recent review of physics at photon colliders can be
found in TESLA/SBLC Conceptual Design Report~\cite{TESLA} and in the
talk of G. Jikia at this workshop.

The resonance production of Higgs in \GG\ collisions and measurement
of its $\gamma\gamma$ width is a task of primary importance.

 Cross sections of the charged particle production in \GG\ collisions
are  higher than those in \EPEM\ collisions. At $E \gg Mc^2$
the ratio of cross sections are the following ($R_{XX}=\sigma_{\GG\to
X^+X^-}/\sigma_{\EPEM\to X^+X^-}$):
$\;R_{H^+H^-}\sim4.5;\; R_{t\bar{t}}\sim4;\;
R_{W^+W^-}(|cos \vartheta|<0.8)\sim15; \;
R_{\mu^+\mu^-}(|cos\vartheta|<0.8)\sim8.5$.

To have the same statistics in \GG\ collisions the luminosity may be
smaller than that in \EPEM\ collisions at least by a factor of 5.
 Note that result in \GG\ and \EPEM\ collisions are complimentary
even for the same final states because diagrams are different (for
example, the vertex $\gamma\gamma$WW can be studied only in \GG\
collisions).

 A reasonable scaling for the required \GG\ luminosity (in the high
energy peak of the luminosity distribution) at \GG\ collider is
\begin{equation}
\LGG\ \sim 3\times10^{33}S(\TEV^2),\;\CMS.
\end{equation}
With such a luminosity  one can detect
$3.5\times10^3\; H^+H^-;\;  2\times10^4 \;
\mu^+\mu^-(|cos\vartheta|<0.8);\;\; 2\times10^4 \; t\bar{t};\;\;
2\times10^5 \; W^+W^-(|cos\vartheta|<0.8);\;\; 2\times10^6\;S(\TEV^2)
\;W^+W^-$ for the time $t=10^7\;c\;$. 
 Somewhat larger luminosity ($\sim 10^{33}$) is required for the
 search and study of the ``intermediate'' ($M_H \sim 100-200\;\GEV$)
 Higgs boson.

 With an electron beam considered in current projects~\cite{NLC}, 
 \cite{TESLA} with 2E$ \sim $500 GeV one can obtain \LGG\ $\sim
  10^{33}\;\CMS$ at $z =W_{\gamma\gamma}/2E>0.7$ (see next section). 
  It is determined only by the
  ``geometric'' \EMEM\ luminosity. Using beams with smaller emittances
  one can get higher luminosity.  Analyses of principle restrictions
  on luminosity of photon colliders have shown~\cite{TSB2} (see sect.4)
  that at $2E \le 5\; \TEV$ one can obtain (in principle) $\LGG \ge
  10^{35}\; \CMS.$
\section{Current projects}
  Recently, two groups NLC~\cite{NLC} and TESLA/SBLC~\cite{TESLA} (and
soon JLC) have published Conceptual Design Reports of their linear
collider projects containing comprehensive appendixes devoted to \GG,
\GE\ options. Below is a short review of these designs.
\subsection{Collision schemes}
Two collision scheme were considered. \underline{scheme A}
(``without deflection''). There is no magnetic deflection of spent
electrons and all particles after the conversion region travel to the
IP. The conversion point may be situated very close to the IP;
\underline{scheme B} (``with deflection''). After the conversion
region, particles pass through a region with a transverse magnetic
field ($B\sim$ 0.5--1 T) where electrons are swept aside. Thereby one
can achieve  more or less pure \GG\ or \GE\ collisions.

In both schemes, the removal of the disrupted spent beam is done by
using the crab crossing scheme with the crossing angle about 30 mrad.
The maximum disruption angle does not exceed 10 mrad and outgoing
beams travel outside the final quads located at a distance about 2 m
from the IP.
\subsection{Conversion region. Requirements to lasers. Optics at the IP.}
The conversion region is situated at the distance $b \sim
1.5\gamma\sigma_y =$ 0.5--1.5 cm from the IP. An optimum laser wave length
for the collider with 2E = 500 GeV is about 1 $\mu m$ (for
$x=4E_0\omega_0/m^2c^4=4.8$) and grows proportionally to the beam
energy. The required flash energy for obtaining the conversion coefficient
$k\sim 0.65$ is about 1--4 J for an electron bunch length
$\sigma_z=$ 0.1--0.5 mm, laser peak power is about 0.5--0.7 TW,
average power is about 20 kW.

\begin{figure}[thb]
\centering
\epsfig{file=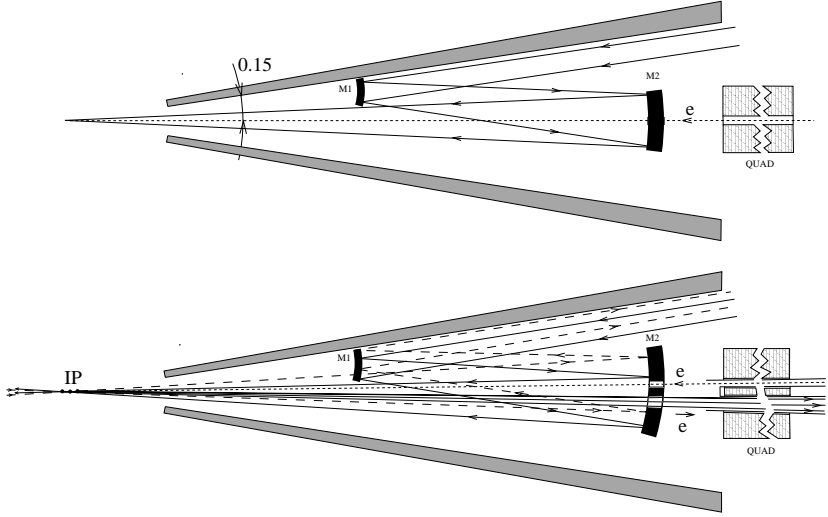,width=4.3in}
\caption{Layout of laser optics near the IP; upper - side view,
down - top view, dashed lines -- exit path of light coming from the
left through one of the CP points (right to the IP), the distance
between the IP and quads is about 2 m}
\label{opti}
\end{figure}

Obtaining such parameters is possible with either solid state lasers or
free electron lasers.  For $\lambda > 1\;\mu m$ (E$_0 >$ 250--300 GeV)
FEL is the only option seen now. The  possible layout of optics near the IP
 is shown in fig~\ref{opti}.

\subsection{Luminosity}

  In current projects, the \GG\ luminosity is determined by the
``geometric'' ee--luminosity. Due to the absence of
beamstrahlung, beams in \GG\ collisions can have much smaller horizontal
beam size than that in \EPEM\ collisions, therefore the beta functions were
taken as small as possible (some restrictions are posed by the Oide
effect connected with chromatic aberrations due to synchrotron
radiation in the final quads). 

   Typical \GG\ luminosity distribution is broad with its peak at maximum
invariant masses at $z=W_{\GG}/2E_0\sim 0.8$ (for $x$=4.8).  The region
$z>0.65$ is the most valuable part of luminosity due to high energy
and high degree of polarization.  The luminosity in this part is
about 10\% of the geometric ee luminosity.

 The results of simulations for different projects are the
   following. For the ``nominal'' beam parameters (the same as in
   \EPEM\ collisions) and the optimum final focus system the
   luminosity $\LGG(z>0.65) \sim (0.8/1.2/0.7)\times10^{33}$ \CMS\ for
   NLC/TESLA/SBLC.  The peak luminosity is also an important
   characteristic, it is  approximately equal to $d\LGG/dz \sim 7
   \LGG(z>0.65)/z_{max}$. These numbers are  close for the schemes
   with and without magnetic deflection.

 In the scheme without deflection, \GE\ collisions can be studied
simultaneously with \GG\ collisions.  In this case the \GE\ luminosity
is even higher than \LGG\ by a factor of 1.5 (this is valid only for
considered beam parameters; for very small beam sizes $L_{\GE} \ll \LGG$
due to beam--beam repulsion). The magnetic deflection allows to obtain almost
clean \GE\ collisions with FWHM$\sim$7\%.

There are several possibilities for increasing  luminosity. 

1) Reduction of the horizontal emittance by optimizing the damping
rings. For example, at the TESLA, the decrease in \ENX\ by a factor
3.5 leads to an increase in \LGG\ up to $3\times 10^{33}$ \CMS.  

2) One can use the low emittance RF-photoguns instead of damping
rings. Unfortunately, even with best photoguns the luminosity will be
somewhat lower than that with damping rings. However, there is one possible
solution. The normalized emittance in photoguns is approximately
proportional to the number of particles in the electron bunch. It seems
possible to merge (using some difference in energies) many ($N_g \sim
5-10$) low current beams with low emittances to one high current beam
with the same transverse emittance. This gives the gain in luminosity more
than by a factor $N_g$ in comparison with one photogun (``more'' because
the lower emittance allows  smaller beta functions due to the Oide effect). 
Joining beams from five photoguns with
experimentally achieved parameters leads at TESLA/SBLC to \LGG\ =
(3--4)$\times 10^{33}$ \CMS.

 For a considerable step in luminosity the beams with much lower
emittances are required that needs the development of new approaches such
as a {\it laser cooling}~\cite{TSB1} (sect.4.2).
Potentially this method allows to attain the geometric luminosity
by two orders higher than those achievable by the methods discussed
above.  One example~\cite{TESLA} of the luminosity distributions for the
``super'' TESLA with round beams and emittances by a factor 50 lower
than that achieved with  RF--photoguns   is shown in fig~\ref{dg5}. Beam
parameters and resulting luminosities are given below.
\begin{figure}[!hbtp]
\centering
\epsfig{file=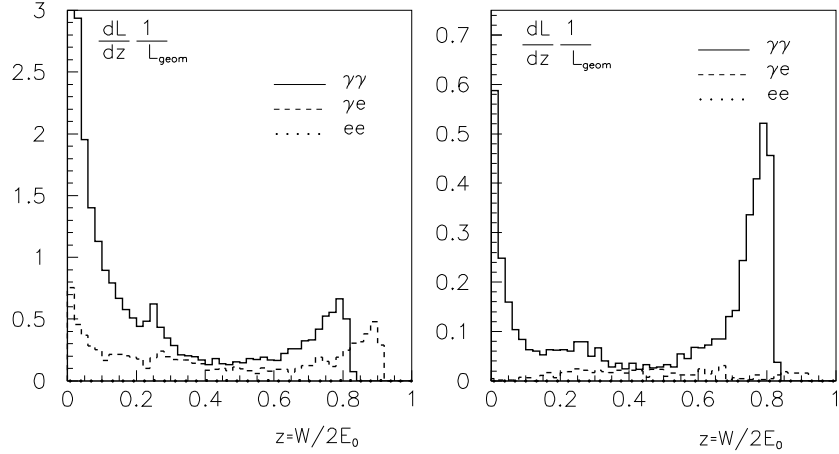,width=4.5in, bb= 50 50 510 280}
\caption{Luminosity spectra for the ``super'' TESLA
parameters (see the text). Left  -- without the deflection;
right -- \GG\ collisions with the magnetic deflection ($B=0.5$ T)} 
\label{dg5}
\end{figure}

{\centering{Electron beam parameters of the ``super'' TESLA}

$N = 3.63 \times 10^{10}$, $\sigma_z = 0.5$~mm, $2E = 500$~GeV, $f = 5.65$~kHz,
$\epsilon_{nx} = \epsilon_{ny} = 0.2 \times 10^{-6}$~m~rad, 
$\beta_x = \beta_y = 0.5$~mm, $\sigma_x=\sigma_y=14$~nm, 
$L_{geom}=2 \times 10^{35}$~cm$^{-2}$~c$^{-1}$

\underline{Luminosities without deflection: $b = \gamma \sigma_y = 0.7$~cm},

$L_{\GG}=1.15 \times 10^{35}$ ,
$L_{\GG} (z>0.65) = 1.5 \times 10^{34}$~cm$^{-2}$~c$^{-1}$,

$L_{\GE}=3.6 \times 10^{34}$,
$L_{\GE} (z>0.65) = 1.2 \times 10^{34}$~cm$^{-2}$~c$^{-1}$,

\underline{Luminosities with magnetic deflection: $b = 1.5$~cm, $B = 0.5$~T},

$L_{\GG}=2 \times 10^{34}$ ,
$L_{\GG} (z>0.65) = 1 \times 10^{34}$~cm$^{-2}$~c$^{-1}$,

$L_{\GE}=2.5 \times 10^{33}$,
$L_{\GE} (z>0.65) = 6 \times 10^{32}$~cm$^{-2}$~c$^{-1}$,

 }

   Results are impressive: \LGG $(z>0.65) =$ (1--1.5)$\times 10^{34}$
\CMS\ (2--3 times higher than those in \EPEM\ collisions). It is not the
limit (see sect.4.3)

\section{New ideas}
\subsection{Laser cooling}

  Recently~\cite{TSB1}, a new method was considered --- laser cooling of
electron beams --- which allows, in principle, to reach $\LGG \geq
10^{35}$ \CMS.

The idea of laser cooling of electron beams is very simple.  During a
collision with optical laser photons (in the case of strong field it
is more appropriate to consider the interaction of an electron with an
electromagnetic wave) the transverse distribution of electrons
($\sigma_i$) remains almost the same. Also, the angular spread
($\sigma_i^{\prime}$) is almost constant, because for
photon energies (a few eV) much lower than the electron beam energy
(several GeV) the scattered photons follow the initial electron
trajectory with a small additional spread. So, the emittance $\EI =
\SI \SIP$ remains almost unchanged. At the same time, the electron
energy decreases from $E_0$ down to $E$. This means that the
transverse normalized emittances have decreased: $ \EN = \gamma \E =
\EN_0(E/E_0)$.  One can reaccelerate the electron beam up to the
initial energy and repeat the procedure. Then after N stages of
cooling $ \EN /\EN _0 = (E/E_0)^N$ (if \EN\ is far from its limit).

 Some possible sets of parameters for the laser cooling are: $E_0 = 4.5$ GeV,
$l_e=0.2 $ mm, $\lambda = 0.5$ \MKM, flash energy $A \sim 10 $ J.  The
final electron bunch will have an energy of 0.45 \GEV\ with an energy
spread $\sigma_E/E \sim 13 \%$, the normalized emittances \ENX,\ENY\
are reduced by a factor 10.  A two stage system with the same parameters
gives 100 times reduction of emittances. The limit on the final
emittance is $\ENX\ \sim\ENY\ \sim2\times 10^{-9}\;$ m~rad at
$\beta_i= 1\; \MM$.  For comparison, in the TESLA (NLC)
project the damping rings have $\ENX\ =14(3)\times 10^{-6}\;$ m~rad,
$\ENY\ =25(3)\times 10^{-8}\;$ m~rad. 

This method requires a laser system even more powerful than that for e $\to
\gamma$ conversion. However, all the requirements are reasonable taking
into account fast progress of laser technique and time plans of linear
colliders. A multiple use of the laser bunch can reduce considerably an average
laser power.

\subsection{Stretching of laser focus depth}

 The laser and electron beams interact with each other most efficiently
when laser and electron beams have the same duration and the depth of
laser focus (Rayleigh length) is somewhat shorter than the beam
length. It turns out that in many cases, the density of laser photons
is so high that instead of the Compton scattering an electron
interacts simultaneously with many photons (synchrotron radiation).
This is not desirable since in the regime of strong field
($eB\lambda>mc^2$) the spectrum of scattered photons after
conversion region is not so peaked as in the  Compton
scattering case. In the method of laser cooling the strong field leads to
higher values of minimum emittance and higher polarization loss.  Of
course, one can take laser bunch longer to keep collision probability
constant and the density of photons below the critical value.
However, in this case, the laser flash energy should be larger than that 
under optimum conditions given in the beginning of
this paragraph.  Due to this nonlinear QED effect  the laser flash
energy required for photon colliders  should grow
proportionally to the collider energy~\cite{TEL90},\cite{TEL95}.

Recently~\cite{TSB1}, it was found how to avoid this problem. In the
suggested scheme the focus depth is stretched without changing the
radius of this area. In this case, the collision probability remains
the same but the maximum value of the field is smaller.  The solution
is based on use of chirped laser pulses and chromaticity of the
focusing system~\footnote{In a chirped pulse the wave length is
linearly depends on longitudinal position. Such pulses are obtained
and used now in all short-pulse lasers.}.  In this scheme, the laser
target consists of many laser focal points (continuously) and light
comes to each point exactly at the moment when the electron bunch is
there.  One can consider that a short electron bunch collides on its
way sequentially with many short light pulses of length $l_{\gamma}
\sim l_e$ and focused with $2Z_R \sim l_e$.

The required flash energy in the scheme with a stretched laser focus is
determined only by diffraction and at the optimum wave length ($x$=4.8)
does not depend on the collider energy. The stretching of laser focus
enables a substantial decrease in flash energy in the method of laser
cooling, to achieve minimum emittances and to conserve polarization
of electron beams.

\subsection{Ultimate luminosity and energy of photon colliders}

  The only collision effect restricting \GG\ luminosity at photon
  colliders is the coherent pair creation which leads to the conversion of
  a high energy photon into \EPEM\ pair in the field of opposing
  electron beam~\cite{CHEN},\cite{TEL90},\cite{TEL95}. There
  are three ways to avoid this effect: a) to use flat beams; b) to
  deflect the electron beam after conversion at a sufficiently large
  distance from the IP; c) under certain conditions (low beam energy,
  long bunches) the beam field at the IP is below the critical one due
  to the repulsion of electron beams \cite{TELSH}.  The problem of
  ultimate luminosities for different beam parameters and energies was
  analyzed recently in ref.~\cite{TSB2} analytically and by
  simulation.  Resume is the following. 

\begin{figure}[!hb]
\centering
\epsfig{file=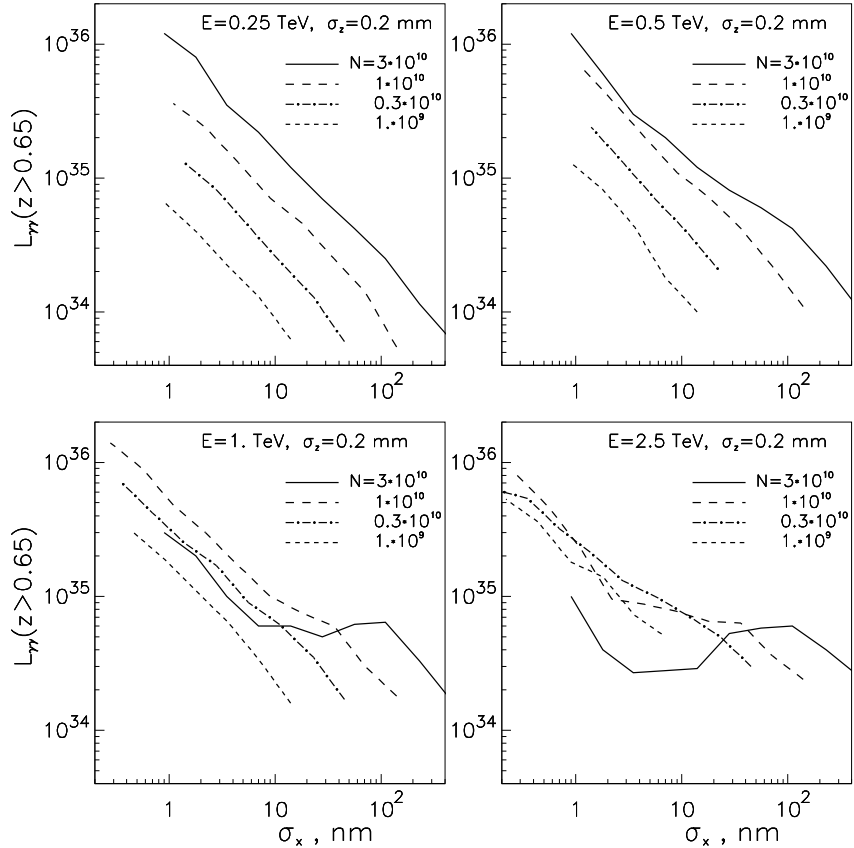,width=11.5cm,bb= 32 50 550 540 }
\caption{Dependence of the \GG\ luminosity on the horizontal beam size
for $\sigma_z = 0.2$ \MM, see comments in the text.}
\label{sb3}
\end{figure}

  The maximum luminosity is attained when the conversion point is
  situated as close as possible to the IP: $b=3\sigma_z+0.04E[\TEV]$
  cm (here the second term is equal to the minimum length of the
  conversion region). In this case, the vertical radius of the photon
  beam at the IP is also minimum: $a_{\gamma}\sim b/\gamma$ (assuming
  that the vertical size of the electron beam is even smaller).  An
  optimum horizontal beam size ($\sigma_x$) depends on the beam
  energy, number of particles in a bunch and bunch length.  The
  dependence of the \GG\ luminosity on $\sigma_x$ for various energies
  and number of particles in a bunch is shown in fig.\ref{sb3}. The
  bunch length is fixed to be equal to 0.2 mm. The collision rate is
  calculated from the total beam power which is equal to 15E[TeV] MW
  (close to that in current projects). In the fig.4 we see that at low
  energies and small number of particles the luminosity curves follows
  their natural behaviour $L \propto 1/\sigma_x$ while at high energy
  and large number of particles in a bunch the curves make zigzag
  which is explained by $\gamma \to$ \EPEM\ conversion in the field of
  the opposing beam.

   What is remarkable in these results? First of all, the maximum
   attainable luminosities are huge. At low energies there are no
   coherent pair creation even for a very small $\sigma_x$ when the
   field in the beam is much higher than the critical field
   $B_{cr}=\alpha e/\gamma r^2_e$. This is explained by the fact that
   beams during the collision are repulsing each other so that the
   field on the beam axis (which affects on high energy photons) is
   below the critical field. It means that the $\gamma\gamma$
   luminosity is simply proportional to the geometric
   electron-electron luminosity (approximately $\LGG(x>0.65)\sim 0.1
   L_{ee}$) for $\sigma_x,\sigma_y > b/\gamma \sim 3\sigma_z/\gamma +
   0.2$ nm.  For the energies 2E$ <$ 2 TeV which are in reach of next
   generation of linear colliders the luminosity limit is much higher
   than it is required by our scaling low given by Eq.1.

\subsection{Backgrounds}

One of important problems at high
luminosities is a background due to relatively large total cross section
$\sigma_{\GG\to hadrons}\sim 5\times10^{-31}\;$cm$^2.$ 

 The average number of hadron events/per bunch crossing is about one
at $\LGG(z>0.65)=2\times 10^{34}\;\CMS$ at the typical collision rate
10 kHz.  However, in the scheme without deflection the total \GG\
luminosity is larger than the ``useful'' $\LGG(z>0.65)$ by a factor
5--10. This low energy collisions increase background by a factor
2--3~\cite{TESLA}.

  Let us assume the photon collider luminosity to be $\LGG(z>0.65)\sim
10^{35}$ \CMS\ (top of our dreams), this leads to about 15
(effectively) high energy $\GG\to hadron$ events per bunch
crossing. Approximately the same number ($\sim$30) of events/collision
is expected in detectors at the LHC . However, there is an important
difference between pp and \GG\ colliders: in the case of an
interesting event (high $P_t$ jets or leptons) the total energy of
final products at photon colliders is equal to $E_{cm}$, while at
proton colliders it is only about (1/6)$E_{cm}$. In comparison with
the pp collider the ratio of the signal to background at photon
colliders is better by a factor of 6 at the same number of hadronic
events per crossing. Note, however, that at NLC and JLC the time
between collisions is only about 1.5 ns and background from a few
neighbouring events will overlap.  At more realistic top \GG\
luminosities about $\LGG(z>0.65)\sim 10^{34}$ \CMS\ even with this
fact the background conditions will be acceptable.  These arguments and
detailed simulation~\cite{TESLA} show that the problem of hadronic
background  is not dramatic for photon colliders.

\section*{Acknowledgments}

I would like to thank the organizers of ``Photon 97'' for the nice
Workshop  which was one of important steps towards photon colliders. 

\end{document}